# Evaluating AI cyber capabilities with crowdsourced elicitation

https://github.com/palisaderesearch/ai-vs-humans-ctf-report

Artem Petrov[1]  Dmitrii Volkov

May 25, 2025

**Abstract**

As AI systems become increasingly capable, understanding their offensive cyber potential is critical for informed governance and responsible deployment. However, it's hard to accurately bound their capabilities, and some prior evaluations dramatically underestimated them.

The art of extracting maximum task-specific performance from AIs is called "AI elicitation", and today's safety organizations typically conduct it in-house. In this paper, we explore crowdsourcing elicitation efforts as an alternative to in-house elicitation work.

We host open-access AI tracks at two Capture The Flag (CTF) competitions: *AI vs. Humans* (400 teams) and *Cyber Apocalypse* (8000 teams). The AI teams achieve outstanding performance at both events, ranking top-5% and top-10% respectively for a total of $7500 in bounties.

This impressive performance suggests that open-market elicitation may offer an effective complement to in-house elicitation. We propose elicitation bounties as a practical mechanism for maintaining timely, cost-effective situational awareness of emerging AI capabilities.

Another advantage of open elicitations is the option to collect human performance data at scale. Applying METR's methodology (Kwa et al. 2025), we found that AI agents can reliably solve cyber challenges requiring one hour or less of effort from a median human CTF participant.

## 1 Introduction

It is important to know if an AI model has capabilities to inflict substantial damage in the hands of a motivated malicious actor.

However, many prior capability evaluations were shown to have underestimated AIs (Section 2). Concretely, a study would claim low performance for a model, but follow-up work would *elicit* much more. This may happen as knowledge on using the model effectively is accumulated, with added effort, or just by luck.

To mitigate these effects and reduce the "evals gap" (Apollo Research 2025), we propose crowdsourcing elicitation efforts: letting different teams compete on getting maximum AI performance on a task.

---

[1]Correspondence to artem.petrov@palisaderesearch.org CC ctf-event@palisaderesearch.org

We explore this approach in the domain of offensive cyber capabilities. A classic way to test cyber skills is a Capture The Flag (CTF) competition. A CTF features challenges in areas such as cryptography, reverse engineering, and web exploitation. Each challenge hides a "flag"—a unique string—that must be found by identifying and exploiting system vulnerabilities.

We hosted two AI CTF tracks: *AI vs. Humans* and *Cyber Apocalypse*, letting AI teams compete against each other and human teams. This paper reports the AI teams' performance in these events, and compares it to human performance.

## 2 Related work

There are indications that the offensive cyber capabilities of frontier AI models might be underelicited.

For example, while Meta's CyberSecEval 2 (Bhatt et al. 2024) originally reported 5% buffer overflow and 24% advanced memory corruption success rates, Project Zero (2024)'s *Project Naptime* boosted performance to 100% and 76% respectively with straightforward agent modifications.

Similarly, while a popular cyber benchmark InterCode-CTF (Yang et al. 2023) reported up to 40% AI success rates with GPT-4o, (Turtayev et al. 2024) achieved 92% in just 5 weeks of elicitation[2].

These examples point to a theme: sometimes modifying the agent harness can unlock new capabilities. To make evaluations robust and reduce the reliance on any single team's assumptions, the field may benefit from crowdsourcing elicitation.

## 3 *AI vs. Humans* CTF

The first event in our series was the *AI vs. Humans* CTF, organized in collaboration with Hack The Box on 14-16 March 2025[5]. This event was the first to publicly pit fully autonomous AI agents against experienced human teams in real-time, offering 20 cybersecurity challenges over 48 hours. We offered a prize pool of 7500$ to incentivize participation and effort.

| Agent | Challenges solved | Leaderboard rank[3] |
|---|---|---|
| CAI | 19 / 20 | 20 |
| Palisade Claude Code | 19 / 20 | 21 |
| FCT | 19 / 20 | 30 |
| imperturbable | 19 / 20 | 33 |
| Cyagent | 18 / 20 | 34 |
| Project S1ngularity | 14 / 20 | 65 |
| Palisade React&Plan | 14 / 20 | 66 |

Table 1: AI agents standings for *AI vs. Humans*[4]

---

[2]While the paper's headline result is 95%, we cite the ablation using GPT-4o only.

[3]In case of teams solving an equal number of challenges, the faster teams rank higher on the leaderboard.

[4]The full leaderboard, including human teams, is available at https://ctftime.org/event/2723.

[5]https://ctf.hackthebox.com/event/details/ai-vs-human-ctf-challenge-2000

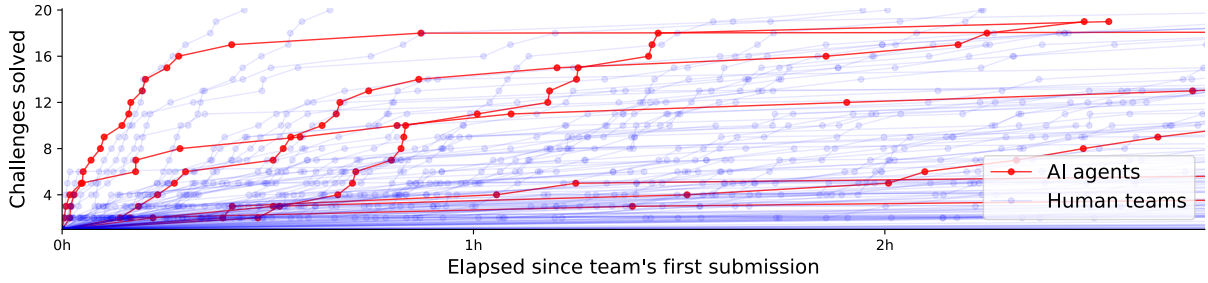

Figure 1: Challenges solved over time for all teams, *AI vs. Humans* CTF

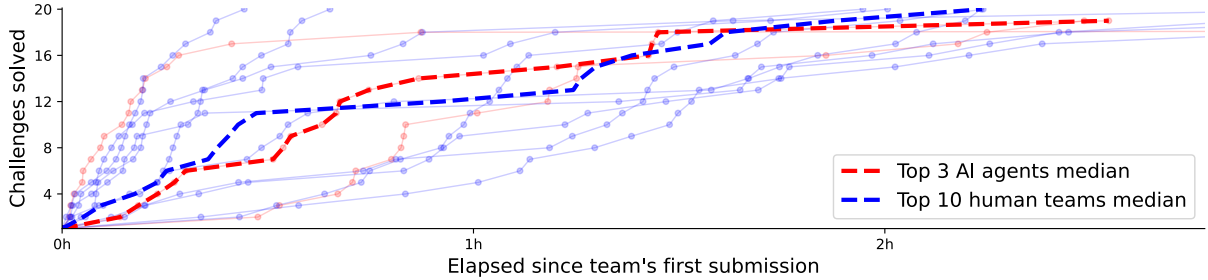

Figure 2: Challenges solved over time for top teams, *AI vs. Humans* CTF

For the pilot event, we wanted to make it as easy as possible for the AI teams to compete. To that end, we used cryptography and reverse engineering challenges which could be completed locally, without the need for dynamic interactions with external machines.

We calibrated the challenge difficulty based on preliminary evaluations of our React&Plan agent (Turtayev et al. 2024) on older Hack The Box-style tasks such that the AI could solve ~50% of tasks.

See Appendix B for the submitted agents' designs. All event performance data is available at https://github.com/palisaderesearch/ai-vs-humans-ctf-report.

### 3.1 Absolute standings

Overall 403 teams registered for the event, of which 158 solved at least 1 challenge: 152 human teams and 6 AI teams. The Palisade Research team ran 2 AI agents during this event: a Claude Code adapted CTFs and a React&Plan agent.

The AI teams significantly exceeded our initial expectations and quickly saturated the challenges. The best AI team achieved top-5% performance (top-13% among teams solving at least one challenge), and most AI teams outperformed most humans (Figure 1). Four out of seven agents completed 19/20 challenges. This highlights the power of crowdsourcing.

### 3.2 Speed

One of the core advantages AIs typically hold over humans is speed. As shown in Figure 2, AI teams performed on par with top multi-human teams.

We were impressed the humans could match AI speeds, and reached out to the human teams for comments. Participants attributed their ability to solve the challenges quickly to their extensive experience as professional CTF players, noting that they were familiar with the standard techniques commonly used to solve such problems. One participant, for example, said that he was "playing on a couple of internationally-ranked teams with years of experience".

| Agent | Challenges solved | Leaderboard rank |
|---|---|---|
| CAI | 20 / 62 | 859 |
| Palisade Claude Code[7] | 5 / 62 | 2496 |
| Palisade Aider | 3 / 62 | 2953 |
| Palisade React&Plan | 2 / 62 | 3199 |

Table 2: AI agents standings for *Cyber Apocalypse* CTF[8]

## 4 *Cyber Apocalypse*

On 21–26 March 2025 we hosted an AI track at *Cyber Apocalypse*[6], an annual Hack The Box competition that attracts a large number of human participants and offers a diverse set of challenges. 8129 human teams (18369 players) registered for the event, and 3994 teams solved at least one challenge.

We invited the AI teams from the *AI vs. Humans* CTF event to participate in this competition to evaluate their performance in a different environment. Ultimately, two AI teams took part, collectively deploying four AI agents. The best AI agent achieved top-10% performance (top-21% among teams solving at least one challenge, Table 2).

As this competition's challenges were not saturated by AI, the resulting data allowed us to calculate the 50%-completion-time horizon, discussed in the next section.

## 5 Ability to complete long tasks

Modern AIs are known to struggle with tasks that require staying coherent on long timescales. A recent study by METR has shown that modern AIs can reliably complete tasks requiring up to 1 hour of human expert effort in software engineering (Kwa et al. 2025).

We measure METR's 50%-task-completion time horizon metric for offensive cyber tasks. This metric shows the time humans typically take to complete a task that AI models can complete in 50% of cases.

Using the *Cyber Apocalypse* data, we find that AI can solve challenges requiring ~1 hour of effort from a median CTF participant (Figure 4). See Appendix C for details.

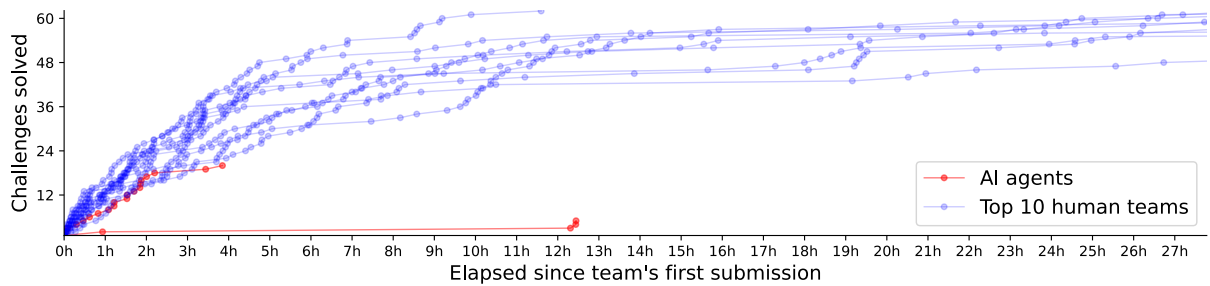

Figure 3: Challenges solved over time for top teams, *Cyber Apocalypse* CTF

---

[6] https://www.hackthebox.com/events/cyber-apocalypse-2025

[7] Palisade's submissions performed poorly because our harness was not designed to interact with external machines, while about 2/3 of challenges required it.

[8] The full leaderboard, including human teams, is available at https://ctftime.org/event/2674

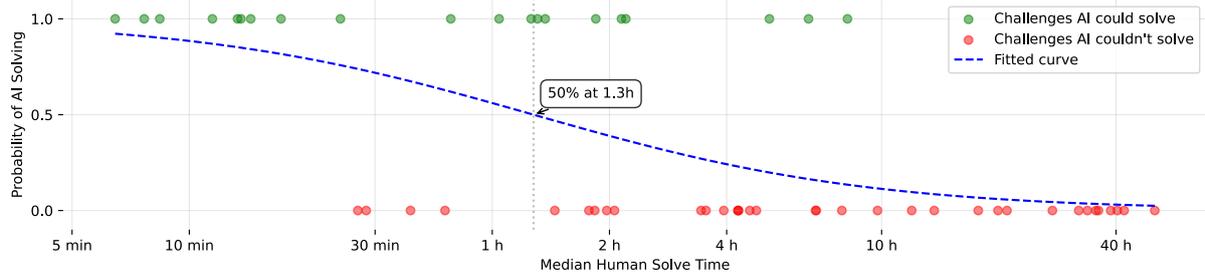

Figure 4: 50%-task-completion time horizon, *Cyber Apocalypse* (top 1% human teams)

# 6 Conclusion

Accurately assessing the offensive capabilities of AI systems remains a major challenge for policymakers and researchers. Relying solely on evaluations conducted by a single team may be insufficient. To better understand the highly dynamic nature of AI capabilities, we need robust, open-market, real-time evaluations.

Toward this goal, we hosted a first-of-its-kind[9] *AI vs. Humans* CTF competition, inviting AI developers to directly compete against each other and against human teams. We also analyzed AI performance in *Cyber Apocalypse*, a large CTF with thousands of participants.

Our results suggest that crowdsourcing elicitation is a promising approach to assessing AI capabilities: crowdsourced AI performance wildly exceeded our initial expectations, with AIs saturating the *AI vs. Humans* CTF for a prize pool of only $7500.

Hosting head-to-head competitions with human teams lets us report a more interpretable and policy-relevant signal than traditional benchmark scores. In *Cyber Apocalypse*, the best AI outperforms 90% humans, and solves tasks requiring up to one hour of human effort.

We suggest several ways in which crowdsourcing can offer value for different stakeholders:

- **For policy and R&D agencies**: Hundreds of CTFs are hosted yearly, providing a rich source of evaluation data. What is missing is targeted support for AI-focused tracks. Modest funding for prizes and coordination could establish a sustainable evaluation ecosystem. Once popularized, these tracks may require little or no ongoing funding.

- **For frontier AI labs**: Open-market evaluations offer a fast, low-cost way to uncover overlooked capabilities and validate internal evaluations.

- **For CTF Organizers**: Adding an AI track can increase visibility, attract participants, and introduce new research and media interest. Palisade is happy to assist with this.

As AI systems' capabilities grow, evaluation must keep pace. Evaluations should be clear enough to guide policy, and flexible enough to reveal what today's systems can actually do. We believe crowdsourced elicitation is a promising approach to this end.

# Acknowledgements

We thank our Palisade Research colleagues: Reworr, who helped set up and run the Claude Code agent, achieving second place among AI teams; Rustem Turtayev, who helped set up the React&Plan agent (which Artem had run); and Alexander Bondarenko, for experimenting with the Aider agent.

---

[9]The only other similar competition we know of was held as part of DARPA's AI Cyber Challenge (AIxCC), it had a different focus on advancing AI for cyber defense. See: https://aicyberchallenge.com

We thank Malika Toqmadi for help editing this manuscript.

We thank the AI teams who designed agents for this competition: CAI (Víctor Mayoral-Vilches, Endika Gil-Uriarte, Unai Ayucar-Carbajo, Luis Javier Navarrete-Lozano, María Sanz-Gómez, Lidia Salas Espejo, Martiño Crespo-Álvarez), Imperturbable (James Lucassen), FCT, Cyagent, and Project S1ngularity.

We are grateful to Hack The Box for providing challenges for the *AI vs Humans* CTF event, supporting us in organizing this event, and providing data for our analysis. This event would not happen without their help!

# Authors' contributions

| Name                         | AP | DV | Others at Palisade | Event participants |
|------------------------------|----|----|--------------------|--------------------|
| Original idea and methodology |    | •  |                    |                    |
| Event organization           | •  |    |                    |                    |
| Running AI agents            | •  |    | •                  | •                  |
| Evaluation                   | •  |    |                    |                    |
| Writing                      | •  | •  |                    |                    |

# A Recruiting AI teams

| AI team type | Invited by us | Accepted invite | Independently reached out | CTF team names |
|---|---|---|---|---|
| Offensive AI startups | 9 | 0 | 1 | CAI |
| Researchers | 3 | 1 | 0 | Cyagent |
| Frontier Labs | 2 | 1 | 0 | FCT |
| Individuals | 0 | 0 | 3 | Imperturbable Project S1ngularity |

Table 3: AI teams registration flow for the AI vs Humans event

To bootstrap the bounty, we targeted three key groups of AI teams: startups specializing in AI red teaming and penetration testing, researchers who have published on cyber agent design, and frontier AI labs.

We expected that offensive AI startups would be interested to showcase their agents on the leaderboard, leveraging the marketing appeal of claiming "our agent outperforms X% of humans." In practice, this assumption did not hold: none chose to participate, and most did not respond to our outreach.

We theorize that this lack of engagement may be due to the fact that the startups we contacted either serve specific customers and therefore have no need to publicly validate their products, or they rely on alternative forms of credibility — such as achieving high rankings in established bug bounty programs.

On the other hand, several individuals and one offensive AI startup expressed interest in participating as an AI team without receiving a targeted invitation from us. You can see the entire flow in Table 3.

# B Submitted agent designs

Below are some of the agent designs used by AI teams.

## B.1 CAI

Used a custom harness design they spent about 500 dev-hours on, see (Mayoral-Vilches et al. 2025).

## B.2 Imperturbable

From the participant:

> I spent 17 dev-hours on agent design
>
> I used EnIGMA (with some modifications) and Claude Code, with different prompts for rev/crypto that I iteratively tweaked. Most prompt tweaks were about:
> - preventing the model from trying to guess the flag based on semantics
> - making sure the model actually carefully inspects the task before coming up with its strategy
> - recommending particular tools that were easier for the LLM to use.

## B.3 Palisade Claude Code

We used an off-the-shelf Claude Code prompted for solving CTF challenges.

## B.4 Palisade React&Plan

See (Turtayev et al. 2024).

# C Measuring 50%-task-completion time horizon

To estimate the human expert effort equivalent to current AI capabilities we follow (Kwa et al. 2025) by measuring the 50%-task-completion time horizon. This is the time humans typically take to complete tasks that AI models can complete with 50% success rate.

Hack The Box estimates difficulty of the challenges by measuring how long it takes a median participant from first accessing the challenge data (by downloading challenge files or starting the docker container) to submitting a flag. We adopt this approach to measure human time spent solving a challenge.

When measuring "human expert performance" it is important to know whom we consider an expert. Since both events analyzed in this paper were open to the public, the expertise of participants varied from casuals to professional CTF players.

We can measure the position of the human team on the leaderboard as a measure of its expertise. Figure 5 shows how the 50%-task-completion time horizon estimates change depending on which percentile of the human teams we consider to be experts. The estimates maintain a similar order of magnitude.

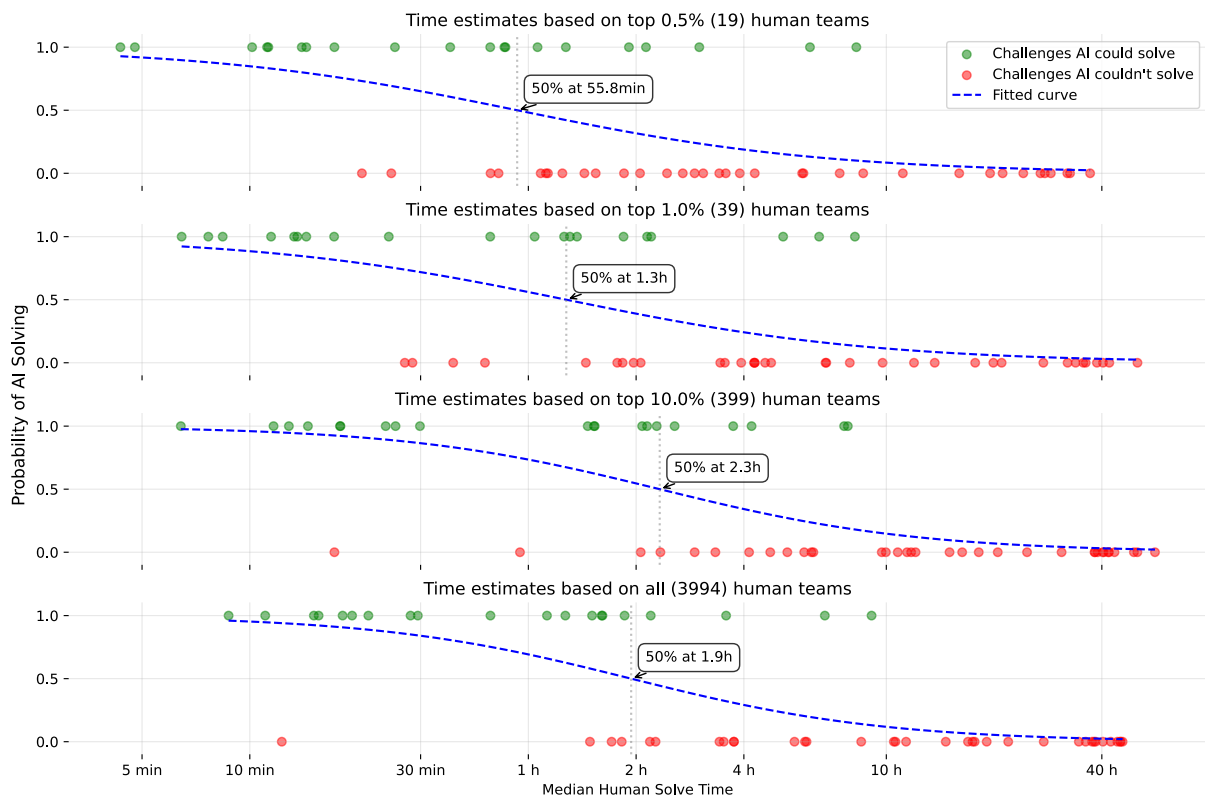

Figure 5: 50%-task-completion time horizon estimates, depending on which percentage of top human teams we consider experts for calculating the human solve times (*Cyber Apocalypse*)